\documentclass[a4paper,oneside,final,notitlepage,onecolumn,12pt]{article}
\usepackage{amsfonts}
\usepackage{epsf}
\usepackage{color}
\usepackage{graphicx}
\usepackage{amssymb,eso-pic}
\usepackage{latexsym}
\usepackage{tabularx}
\usepackage{amsxtra}
\usepackage{hyperref}
\usepackage{t1enc}
\usepackage{amsmath}
\usepackage[T1]{fontenc}
\usepackage{amsmath,accents}
\usepackage{authblk}

\usepackage{framed}
\usepackage{enumerate} 
\usepackage{eufrak}
\usepackage{bm}
\usepackage{mathrsfs}
\usepackage[dvips]{epsfig}
\usepackage{amsthm}
\usepackage{latexsym}
\usepackage{tikz}
\usepackage{mathtools}
\DeclareGraphicsExtensions{.jpg,.jpeg,.pdf,.png,.eps} 

\usepackage{array,amsmath,amssymb}

\usepackage[normalem]{ulem}
\usepackage{color}
\definecolor{blue}{rgb}{0,0,1}
\definecolor{red}{rgb}{1,0,0}

\DeclareFontFamily{OT1}{rsfs}{} \DeclareFontShape{OT1}{rsfs}{m}{n}{
	<-7> rsfs5 <7-10> rsfs7 <10-> rsfs10}{}
\DeclareMathAlphabet{\mycal}{OT1}{rsfs}{m}{n}

\definecolor{DGREEN}{rgb}{0,0.65,0.65}
\definecolor{DCYAN}{rgb}{0,0.45,0.45}
\definecolor{grey1}{rgb}{0.52, 0.52, 0.51}

\definecolor{mybg}{rgb}{212,242,250}

\DeclareFontFamily{OT1}{rsfs}{} \DeclareFontShape{OT1}{rsfs}{m}{n}{
	<-7> rsfs5 <7-10> rsfs7 <10-> rsfs10}{}
\DeclareMathAlphabet{\mathscr}{OT1}{rsfs}{m}{n}

\def\sc{{\hskip 3.5pt {{}^{{}^{{}_{{}_{\bowtie}}}}} \kern -8.pt{}}}  
\def\SC{{\hskip 3.5pt {{}^{{}^{{}^{{}_{{}_{\bowtie}}}}}} \kern -10.5pt{}}}

\usepackage[normalem]{ulem}
\usepackage{color}

\newcounter{mnotecount}

\newcommand{\mnotex}[1]
{\protect{\stepcounter{mnotecount}}$^{\mbox{\footnotesize $\bullet$\themnotecount}}$ 
	\marginpar{\color{red}
		\raggedright\tiny\em
		$\!\!\!\!\!\!\,\bullet$\themnotecount: #1} }

\begin{document}
	
	\title{\textbf{On the evolutionary form of the constraints in electrodynamics}}
	
	\author[,1,2]{Istv\'an R\'acz \footnote{E-mail address:{\tt \,racz.istvan@wigner.mta.hu}}}
	
	\affil[1]{Faculty of Physics, University of Warsaw, Ludwika Pasteura 5, 02-093 Warsaw, Poland}
	
	\affil[2]{Wigner RCP, H-1121 Budapest, Konkoly Thege Mikl\'{o}s \'{u}t  29-33, Hungary}

	\maketitle
	
	\begin{abstract}
		
		The constraint equations in Maxwell theory are investigated. In analogy with some recent results on the constraints of general relativity it is shown, regardless of the signature and dimension of the ambient space, that the ``divergence of a vector field'' type constraints can always be put into linear first order hyperbolic form for which global existence and uniqueness of solutions to an initial-boundary value problem is guaranteed. 
		
	\end{abstract}
	

\section{Introduction}
\parskip 4pt

The Maxwell equations, as we know them since the seminal addition of Ampere's law by Maxwell in 1865, are \cite{Jackson-1999} 
\begin{align}
{\boldsymbol\nabla} \times {\bf H} = {\bf J}  + \partial_t {\bf D}	& \qquad\qquad {\boldsymbol\nabla} \times {\bf E} + \partial_t {\bf B} = 0 \label{eq: M1} \\
{\boldsymbol\nabla} \cdot {\bf D} =  q	& \qquad\qquad\qquad\quad\    {\boldsymbol\nabla} \cdot {\bf B} = 0 \label{eq: M2} \,,
\end{align}
where $\bf E$ and $\bf B$ are the macroscopic electric and magnetic field variables, which in vacuum are related to $\bf D$ and $\bf H$ by the relations ${\bf D}={\boldsymbol\epsilon}_0\, \bf E$ and ${\bf H}={\boldsymbol\mu}_0^{-1}\, \bf B$, where ${\boldsymbol\epsilon}_0$ and ${\boldsymbol\mu}_0$ are the dielectric constant and magnetic permeability, and where $q$ and ${\bf J}$ stand for charge and current densities, respectively. 

The top two equations in \eqref{eq: M1} express that the time dependent magnetic field induces an electric field and also that the changing electric field induces a magnetic field even if there are no electric currents. Obviously there have been plenty of brilliant theoretical, experimental and technological developments based on the use of these equations. Nevertheless, from time to time some new developments (for a recent examples see, for instance, \cite{Deckert-2016,Medina-Stephany-2018}) have stimulated reconsideration of claims which previously were treated as text-book material in Maxwell theory.   

In this short note the pair of simple constraint equations on the bottom line in \eqref{eq: M2} are the centre of interest. These relations for the
divergence of a vector field are customarily treated as elliptic equations. The main purpose of this letter is to show that by choosing basic variables in a geometrically preferred way the constraints in \eqref{eq: M2} can also be solved as evolutionary equations. This also happens in the more complicated
case of the constraints in general relativity \cite{racz_constraints}.   

Once the Maxwell equations \eqref{eq: M1} and \eqref{eq: M2} are given it is needless to explain in details what is meant to be the ambient spacetime (tacitly it is assumed to be the Minkowski spacetime) or the initial data surface (usually chosen to be a ``$t=const$'' hypersurface in Minkowski spacetime). As seen below the entire argument, outlined in more details in the succeeding sections, is very simple. In addition, it applies with almost no cost to a generic ambient space $(M,g_{ab})$, with a generic three-dimensional
initial data surface $\Sigma$. We shall treat the generic case, i.e.~solve the ``divergence of a vector field type constraints'', 
\begin{equation}\label{eq: div-type-eqs}
{\boldsymbol\nabla} \cdot {\bf L} =\ell \, , \quad \textrm{(in index notation)}\quad	D_i L^i = \ell \, ,
\end{equation}
for a vector field ${\bf L}$ or (in index notation) $L^i$ with a generic source $\ell$, on a fixed but otherwise arbitrary initial data surface, $\Sigma$. As an initial data surface can always be viewed as a time slice in an ambient spacetime, $(M,g_{ab})$, it is also straightforward to assign a Riemannian metric $h_{ij}$ to $\Sigma$, the one induced by $g_{ab}$ on $\Sigma$. In \eqref{eq: div-type-eqs} above, $D_i$ stands for the unique torsion free covariant derivative operator that is compatible with metric $h_{ij}$.

Note that for the Maxwell system, given by \eqref{eq: M1} and \eqref{eq: M2}, the two
divergence of a vector field constraints decouple so it suffices to solve them
independently.  Note also that it is easy to see that all the arguments presented in the succeeding subsections generalize to an arbitrary $n\geq 3$ dimension of $\Sigma$.  Nevertheless, for the sake of simplicity, our consideration here will be restricted to the case of three-dimensional initial data surfaces. 

Since the constraints are almost exclusively referred to as elliptic equations
in text-books, one may question the point of putting them into evolutionary form. We believe that the appearance of time evolution in a Riemannian space could itself be of interest on its own right. Nevertheless, it is important to emphasize that there are valuable applications of the proposed new method. For instance, it may offer solutions to problems which are hard to solve properly in the standard elliptic approach. An immediate example of this sort arises in the initialization of the
time evolution of point charges governed by the coupled Maxwell-Lorentz equations. As pointed out recently in \cite{Deckert-2016}, unless suitable additional conditions are applied in addition to the Maxwell constraints, the electromagnetic field develops singularities along the light cones emanating from the original positions of
the point charges.
It is important to be mentioned here that analogous problems arise in the context of the initialization of the time evolution of binary black hole configurations. In both cases singularities are involved which in case of the Maxwell-Lorentz system are located  at the point charges, whereas in binary black hole case at the spacetime singularities. The main task is to construct physically adequate initial data specifications such that they are regular everywhere apart from these singularities. In case of binary black hole configurations this can be done by using the superposed Kerr-Schild metric, as an auxiliary ingredient in determining the freely specifiable fields. 
Then suitable ``initial data'' is chosen, in the distant radiation dominated region, to the evolutionary form of the Hamiltonian and momentum constraints of general relativity. The desired initial data is completed finally by solving the corresponding initial value problem \cite{racz_bh_ID}. A completely analogous procedure is proposed to be used in  initializing the time evolution of a pair of interacting point charges in Maxwell theory. [A detailed outline of this proposal is given in Section \ref{sec: example}.] In this case the ``superposed'' Li\'enard-Wiechert vector potentials are used, as an auxiliary ingredient to prescribe the freely specifiable fields. In addition, suitable initial data has to be chosen to the evolutionary forms of the constraints [see eq.~\eqref{eq: 2new_div-type-eqs} below] in a distant radiation dominated region. The desired initial data can then be completed by solving the evolutionary form of the constrains \eqref{eq: M2} as an initial value problem. 
It is remarkable that while in the conventional elliptic approach some assumptions (in most cases tacit ones) are always used concerning the blow up rate (while approaching the singularities) of the constrained fields, no such fictitious ``inner boundary condition'' is applied anywhere in the proposed new method. It is indeed the evolutionary form of constraints itself that tells to the constrained variables how they should evolve from their weak field values towards and up to the singularities.

An additional, and not the least important, potential advantage of the proposed new method is that it offers an unprecedented flexibility in solving the constraint equations. This originates from the fact that neither 
the choice of the underlying foliations of the three-dimensional initial data surface $\Sigma$ nor the choice of
the evolutionary flow have any limitations. This makes the proposed method applicable to a high variety of problems
that might benefit from this new approach to solving
the constraints.

Another advantage of this new approach
constraints is that, regardless of the choice of foliation
and flow, the geometrically preferred set of variables constructed in carrying out the main steps of the
procedure always satisfy a {\it linear first order symmetric hyperbolic equation}. 
Considering the robustness of the approach,
it is remarkable that, starting with the ``divergence of a vector field constraint'', the global existence of a unique smooth solution for the geometrically preferred dependent variables (under suitable regularity conditions on the coefficients and source terms) is guaranteed for the linear first order symmetric hyperbolic equation (see, e.g.\, subsection VIII.12.1 in \cite{Choquet-2015}).

\section{Preliminaries}\label{sec: prelim}

The construction starts by choosing a three-dimensional initial data surface $\Sigma$ with an induced Riemannian
metric $h_{ij}$ and its associated torsion free covariant derivative operator $D_i$. $\Sigma$ may be assumed to lie in an ambient space $(M,g_{ab})$ whose metric could have either Lorentzian or Euclidean signature. More importantly, $\Sigma$ will be assumed to be a topological product 
\begin{equation}
\Sigma \approx \mathbb{R}\times \mathscr{S}\,,
\end{equation}
where $\mathscr{S}$ could be of a two-surface with arbitrary topology. In the simplest practical case, however,
$\mathscr{S}$ would have either planar, cylindrical, toroidal or spherical topology. In these cases, we may assume that there exists a smooth real function $\rho: \Sigma \rightarrow \mathbb{R}$ whose $\rho=const$ level sets give the $\mathscr{S}_\rho$ leaves of the foliation and
that its gradient ${\partial}_i \rho$ does not vanish,
apart from some isolated locations where the foliation may degenerate.\,\footnote{If, for instance, $\Sigma$ has the topology $\mathbb{R}^3$, $\mathbb{S}^3$, $\mathbb{S}^2\times \mathbb{R}$ or $\mathbb{S}^2\times \mathbb{S}^1$ and it is foliated by topological two-spheres then there exists one, two or, in the later two cases, no points of degeneracy at all. If point charges are involved it may be preferable to place the associated physical singularities at the location of these degeneracy. 
	Note also that we often shorthand partial derivatives $\partial/\partial x^i$ by $\partial_i$.}

\begin{figure}[ht]
	\centerline{\includegraphics[width=0.7\textwidth, angle=0]{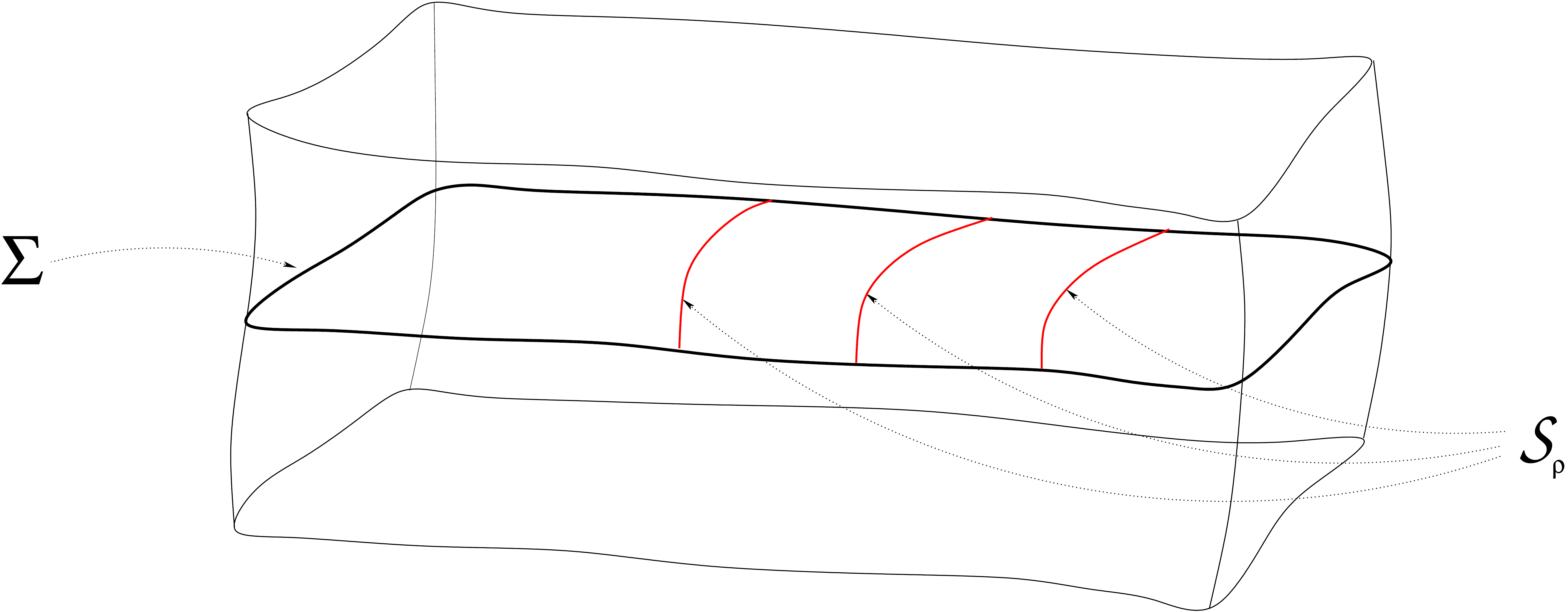}}
	\caption{\footnotesize \label{fig: Fig}  The initial data surface $\Sigma$ foliated by a one-parameter family of two-surfaces $\mathscr{S}_\rho$ is indicated.} 
\end{figure}

The above condition guarantees (as indicated in Fig.\ref{fig: Fig}) that locally $\Sigma$ is smoothly foliated by a one-parameter family of $\rho=const$ level two-surfaces
$\mathscr{S}_\rho$ . 
Given these leaves, the non-vanishing gradient ${\partial}_i \rho$ can be normalized to a unit normal $\widehat n_i = {\partial}_i \rho/\sqrt{h^{ij}({\partial}_i\rho)({\partial}_j\rho})$,
using the Riemannian metric $h_{ij}$.
Raising the index according to $\widehat n^i = h^{ij}\widehat n_j$ gives the unit vector field normal
to $\mathscr{S}_\rho$. The operator $\widehat \gamma{}^i{}_j$
formed form
the combination of $\widehat n_i$ and $\widehat n^i$ and the identity operator $\delta {}^i{}_j$, 
\begin{equation}
\widehat \gamma{}^i{}_j=\delta {}^i{}_j-\widehat n{}^i\widehat n_j
\end{equation}
projects fields on $\Sigma$ to the tangent space of the $\mathscr{S}_\rho$ leaves.

We also apply flows interrelating the fields defined on the successive $\mathscr{S}_\rho$ leaves. A  vector field $\rho^i$ on $\Sigma$ is called a flow if its integral curves intersect each of the leaves precisely once and it is normalized such that $\rho^i \partial_i \rho=1$ holds everywhere on $\Sigma$. The contraction $\widehat N= \rho^j \widehat n_j$ of $\rho^i$ with $\widehat n_i$ and its projection $\widehat N^i=\widehat \gamma{}^i{}_j\,\rho^j$ of $\rho^i$ to the leaves are referred to as the ``lapse'' and ``shift'' of the flow  and we have
\begin{equation}
\rho^i=\widehat{N}\,\widehat n^i+{\widehat N}{}^i\,.
\end{equation}

The inner geometry of the {$\mathscr{S}_\rho$} leaves can be characterized by the metric 
\begin{equation}\label{eq: induced-metric}
\widehat \gamma{}_{ij}=\widehat \gamma{}^{\,k}{}_i\,\widehat \gamma{}^{\,l}{}_j\, h_{kl}
\end{equation}
induced on the $\rho=const$ level surfaces. It is also known that a unique torsion free covariant derivative operator ${\widehat D}_i$ associated with the metric
$\widehat \gamma{}_{ij}$ acts on fields intrinsic to the {$\mathscr{S}_\rho$} leaves, e.g.~acting on the
field ${\rm\bf N}{}_l=\widehat\gamma{}^p{}_l  \,N_p$ obtained by the projection of $N_p$ according to
\begin{equation}\label{eq: cov-der}
{\widehat D}_i {\rm\bf N}{}_j = \widehat\gamma{}^k{}_i \, \widehat\gamma{}^l{}_j  D_k \left[\, \widehat\gamma{}^p{}_l  \,N_p\,\right]\,.
\end{equation}
It is straightforward to check that ${\widehat D}_i$ is indeed metric compatible in the sense that ${\widehat D}_k\widehat \gamma{}_{ij}$ vanishes.

Note also that the exterior geometry of the {$\mathscr{S}_\rho$} leaves can be characterized by the 
extrinsic curvature tensor ${\widehat K}{}_{ij}$ and the acceleration $\dot {\widehat n}_i$ of the unit normal,
given by
\begin{equation}\label{eq: ext-acc}
{\widehat K}{}_{ij}= \frac12\,\mathscr{L}_{\widehat n}  \widehat \gamma{}_{ij} \quad  \textrm{and} \quad
\dot {\widehat n}_i={\widehat n}^l D_l {\widehat n}_i=-{\widehat D}_i \ln {\widehat N}\,,
\end{equation}
where $\mathscr{L}_{\widehat n}$ is the Lie derivative operator with respect to the vector field $\widehat{n}{}^i$
and $ {\widehat N}$ is the lapse of the flow.

\section{The evolutionary form the constraints}\label{sec: main}

This section is to put the divergence type constraint \eqref{eq: div-type-eqs} into evolutionary form. This is achieved by applying a $2+1$ decomposition where, as we see below, the main conclusion is completely insensitive to the choice of the foliation and of the flow. 

Consider first an arbitrary co-vector field $L_i$ on $\Sigma$. By making use of the projector $\widehat \gamma{}^{\,i}{}_j$ defined in the previous section we obtain
\begin{equation}
L_i={\delta^{\,j}}_i\,L_j=(\widehat\gamma{}^{\,j}{}_i + \widehat n{}^j\widehat n_i)\,L_j=\boldsymbol\lambda\,\widehat n_i + {\rm\bf L}{}_i\,,
\end{equation}
where the boldfaced variables $\boldsymbol\lambda$ and ${\rm\bf L}{}_i$ are fields intrinsic to the individual $\mathscr{S}_\rho$ leaves of the foliation of  $\Sigma$. They are defined via the contractions 
\begin{equation}
\boldsymbol\lambda= \widehat  n^l\,L_l \ \ \ {\rm and} \ \ \  {\rm\bf L}{}_i=\widehat\gamma{}^{\,j}{}_i \,L_j\,.
\end{equation}

By applying an analogous decomposition of $D_i L_j$ we obtain
\begin{equation}
D_i L_j= {\delta^{\,k}}_i{\delta^{\,l}}_j D_k\left[\, {\delta^{\,p}}_l \,L_p \,\right] =  (\widehat\gamma{}^{\,k}{}_i + \widehat n{}^k\widehat n_i) (\widehat\gamma{}^{\,l}{}_j + \widehat n{}^l\widehat n_j) D_k \left[\, (\widehat\gamma{}^{\,p}{}_l + \widehat n{}^p\widehat n_l) \,L_p\,\right]\,,
\end{equation}
which, in terms of the induced metric \eqref{eq: induced-metric}, the associated covariant derivative operator, the extrinsic curvature and the acceleration \eqref{eq: ext-acc}, can be written as 
\begin{align}
D_i L_j= \big[\,\widehat D_i \boldsymbol\lambda + \widehat n_i\,\mathscr{L}_{\widehat n}\boldsymbol\lambda\,\big]\widehat n_j + \boldsymbol\lambda\,(\widehat K_{ij}+ & \,\widehat n_i\dot {\widehat n}_j)+  \widehat D_i {\rm\bf L}{}_j - \widehat n_i \widehat n_j \,(\dot {\widehat  n}^l{\rm\bf L}{}_l) \nonumber \\ + {}  & \ \big\{\widehat n_i\,\mathscr{L}_{\widehat n} {\rm\bf L}{}_j - \widehat n_i\,{\rm\bf L}{}_l {\widehat K^l}{}_j - \widehat n_j\,{\rm\bf L}{}_l {\widehat K^l}{}_i   \big\} \,.
\end{align}
By contracting the last equation with the inverse $h^{ij}=\widehat \gamma{}^{\,ij}+\widehat  n^i \widehat n^j$ of the three-metric $h_{ij}$ on $\Sigma$, we obtain
\begin{equation}
D^l L_l= h^{ij} D_i L_j = (\widehat \gamma^{\,ij}+\widehat  n^i \widehat n^j)\,D_i L_j =\mathscr{L}_{\widehat n}\boldsymbol\lambda+\boldsymbol\lambda\,({\widehat K^l}{}_l) +\widehat D^l {\rm\bf L}{}_l+\dot {\widehat n}^l{\rm\bf L}{}_l\,.
\end{equation}
In virtue of \eqref{eq: div-type-eqs} and in accord with the
last equation, it is straightforward to see that the divergence of a vector field constraint can be put into the form
\begin{equation}\label{eq: new_div-type-eqs}
\mathscr{L}_{\widehat n}\boldsymbol\lambda+\boldsymbol\lambda\,({\widehat K^l}{}_l) +\widehat D^l {\rm\bf L}{}_l+\dot {\widehat n}^l{\rm\bf L}{}_l =\ell\,.
\end{equation} 

Now, by choosing arbitrary coordinates $(x^2,x^3)$ on
the $\rho=const$ leaves and by Lie dragging them along the  chosen flow $\rho^i$,  coordinates $(\rho,x^2,x^3)$ adapted to both the foliation $\mathscr{S}_\rho$ and the flow
$\rho^i=(\partial_\rho)^i$ can be introduced on $\Sigma$. In these coordinates, \eqref{eq: new_div-type-eqs} takes the strikingly simple form in terms of the lapse and
shift of the flow, 
\begin{equation}\label{eq: 2new_div-type-eqs}
\partial_\rho \boldsymbol\lambda - \widehat N^K \partial_K \boldsymbol\lambda + \boldsymbol\lambda\,\widehat N\,({\widehat K^L}{}_L) +\widehat N\big[{\widehat D_L {\rm\bf L}{}^L+\dot {\widehat n}_L{\rm\bf L}{}^L }\big]=\ell\,.
\end{equation}

Some remarks are now in order. 
First, \eqref{eq: 2new_div-type-eqs} is a scalar equation whereby it is natural to view it as an equation for the scalar part $\boldsymbol\lambda={\widehat n}_i L^i$ of the vector field $L^i$ on $\Sigma$ and to solve it for  $\boldsymbol\lambda$. All the coefficients and source terms in \eqref{eq: 2new_div-type-eqs} are determined explicitly by freely
specifying the fields ${\rm\bf L}{}^L$ and $\ell$,
whereas the metric $h_{ij}$ and its decomposition in terms of the variables $\widehat N,\widehat N^I,\widehat \gamma_{IJ}$, is also known throughout $\Sigma$.
Thus \eqref{eq: 2new_div-type-eqs} can be solved for $\boldsymbol\lambda$. Note that \eqref{eq: 2new_div-type-eqs}, is manifestly independent of the choice made for the foliation and flow, and also that \eqref{eq: 2new_div-type-eqs} is always a linear hyperbolic equation for $\boldsymbol\lambda$, with $\rho$ ``playing the role of time''.  

\section{A simple example}\label{sec: example}

Though the results in the previous section are mathematically all robust it would be pointless to have the proposed evolutionary form of the constraints unless one could apply it in solving certain problems of physical interest. In order to get some hints how the proposed techniques work, this section is to give an outline of a construction that could be used to get meaningful initialization of the time evolution of a pair of moving point charges in Maxwell theory. 

Recall first that accelerated charges are know to emit electromagnetic radiation. An interesting particular case is when the radiation is emitted by a pair of point charges moving as dictated by their mutual electromagnetic field. To start off choose the $t=0$ time slice in a background Minkowski spacetime. This time slice itself is a three-dimensional Euclidean space $\mathbb{R}^3$ that can be endowed with the conventional Cartesian coordinates $(x,y,z)$ as a three-parameter family of inertial observers has already been chosen in the ambient Minkowski background. 
Assume that on this time slice the two point charges are located on the $y=z=0$ line at $x=\pm a$ (with some $a>0$) each moving with some initial speed. Choose then a one-parameter family of confocal rotational symmetric ellipsoids
\begin{align}
\begin{split}
x &= a \cdot \cosh \rho \cdot \cos\chi  \\
y &= a \cdot \sinh \rho \cdot \sin\chi \cdot \cos \varphi  \\
z &= a \cdot \sinh \rho \cdot \sin\chi \cdot \sin \varphi\,. 
\end{split}
\end{align}
It is straightforward to check that $\mathbb{R}^3$ gets to be foliated by the $\rho=const$ level surfaces, that are confocal rotational ellipsoids
\begin{equation}
\frac{x^2}{a^2\cdot \cosh^2\!\rho}+ \frac{y^2+z^2}{a^2\cdot \sinh^2\!\rho}=1\,,
\end{equation}
with focal points $f_+=(a,0,0)$ and $f_-=(-a,0,0)$. Note also that each member of the two-parameter family of curves determined by the relations $\chi=const,\varphi=const$, with $0<\chi\leq 2\pi$, parameterized by $\rho\,(\geq0)$, intersect $\rho=const$ level surfaces precisely once. The introduced new coordinates $(\rho,\chi,\varphi)$ cover the complement of the two focal points in $\mathbb{R}^3$. Choose this complement as our initial data surface $\Sigma$. These coordinates, adopted to the $\rho: \Sigma \rightarrow \mathbb{R}$ foliation and to the flow vector field $\rho^i$ on $\Sigma$, are such that $\rho^i$ is parallel to the $\chi=const,\varphi=const$ coordinate lines and is normalizes such that $\rho^i(\partial_i\rho)=1$. The pertinent laps and shift, $\widehat N$ and $\widehat N{}^i$, of this coordinate bases vector $\rho^i=(\partial_\rho)^i$ can also be determined as described in Section \ref{sec: prelim}. 

Following then a strategy analogous to the one applied in getting the binary black hole initial data in general relativity \cite{racz_bh_ID} one may proceed as follows. By superposing the Li\'enard-Wiechert vector potentials relevant for the individual point charges, moving with certain initial speeds, determine first the corresponding auxiliary Faraday tensor ${}^{(aux)}F_{ab}$. Restrict it to the $t=0$ initial data surface and extract there the auxiliary electric ${}^{(aux)}{\bf E}$ and magnetic ${}^{(aux)}{\bf B}$ fields. These electric and magnetic parts of ${}^{(aux)}F_{ab}$ are meant to be defined with respect to the aforementioned three-parameter family of static observers [moving in the background Minkowski spacetime with four velocity $u^a=(\partial_t)^a$]. Split these vector fields, as described at the beginning of Section \ref{sec: main}, into scalar and  two-dimensional vector parts we get ${}^{({aux})}E_i={}^{(aux)}\boldsymbol\varepsilon\,\widehat n_i + {}^{(aux)}{\bf\mathcal{E}}{}_i$ and ${}^{(aux)}B_i={}^{(aux)}\boldsymbol\beta\,\widehat n_i + {}^{(aux)}{\bf \mathcal{B}}{}_i$, respectively. The two-dimensional vector parts ${}^{(aux)}{\bf \mathcal{E}}{}_A$ and ${}^{(aux)}{\bf \mathcal{B}}{}_A$, of the auxiliary electric ${}^{(aux)}{\bf E}$ and magnetic ${}^{(aux)}{\bf B}$ fields, are well-defined smooth fields on $\Sigma$. As they encode important information about the momentary kinematical content of the considered system, e.g.~the initial speeds and locations of the involved point charges, the fields ${}^{(aux)}{\bf \mathcal{E}}{}_A$ and ${}^{(aux)}{\bf \mathcal{B}}{}_A$ are used as the freely specified part of data throughout $\Sigma$. Once this has been done the radiation content of the initial data, for the physical ${\bf E}$ and ${\bf B}$, in the far zone has to be introduced by choosing---based on measurements, expectations and/or intuition---two smooth functions, ${}_{(0)}\boldsymbol\varepsilon$ and ${}_{(0)}\boldsymbol\beta$, on a level surface $\rho=\rho_0$ (for some sufficiently large real value of $\rho_0$) in $\Sigma$. These are the initial data to the pertinent forms of \eqref{eq: 2new_div-type-eqs} that can be deduced---as described in Sections \ref{sec: prelim} and \ref{sec: main}---from the constraints equations in \eqref{eq: M2}. 

Remarkably, for arbitrarily small values of $\epsilon>0$ unique smooth solutions $\boldsymbol\varepsilon$ and $\boldsymbol\beta$ to the (decoupled) evolutionary form of the constraint equations exist in the region bounded by the $\rho=\rho_0$ and $\rho=\epsilon$ level surfaces. [One could integrate the equations also outwards, with respect to $\rho=\rho_0$, nevertheless, if one is interested in the behavior of the initial data in the near zone region then the aforementioned domain is the relevant one.] The corresponding unique smooth solutions smoothly extend onto $\Sigma$, even in the $\epsilon \rightarrow 0$ limit, in spite of the fact the solutions are known to blow up at the focal points where the point charges are located initially. Using the unique smooth solution $\boldsymbol\varepsilon$ and $\boldsymbol\beta$, corresponding to the choices made for the initial data ${}_{(0)}\boldsymbol\varepsilon$ and ${}_{(0)}\boldsymbol\beta$ at $\rho=\rho_0$, the initialization of the physical electric and magnetic fields given as $E_i=\boldsymbol\varepsilon\,\widehat n_i + {}^{(aux)}{\bf\mathcal{E}}{}_i$ and $B_i=\boldsymbol\beta\,\widehat n_i + {}^{(aux)}{\bf \mathcal{B}}{}_i$, respectively. Note that they will differ from 
${}^{({aux})}E_i$ and ${}^{({aux})}B_i$ as the initial data ${}_{(0)}\boldsymbol\varepsilon$ and ${}_{(0)}\boldsymbol\beta$, for the pertinent forms of \eqref{eq: 2new_div-type-eqs}, were chosen to differ from ${}^{(aux)}\boldsymbol\varepsilon$ and ${}^{(aux)}\boldsymbol\beta$.
More importantly, once the electric and magnetic fields ${\bf E}$ and ${\bf B}$ get to be  initialized the way prescribed above, the conventional time evolution equations of the coupled Maxwell-Lorentz system [including the two ones in \eqref{eq: M1}] relevant for the pair of interacting point charges should be solved (possible by numerical means). Notably, due to the above outlined initialization, the radiation that will emerge from the consecutive accelerating motion of the pair of point charges is guaranteed to be consistent with the radiation imposed, by specifying initial data ${}_{(0)}\boldsymbol\varepsilon$ and ${}_{(0)}\boldsymbol\beta$, at the $\rho=\rho_0$ level surface located in the far zone.  

\section{Final remarks} 

In virtue of the main result of this note the ``divergence of a vector type constraints'' can always be solved as a linear first order hyperbolic  equation for the scalar part of the vector variable under considerations. As it was emphasized in the introduction robust mathematical results guarantee the global existence of unique smooth solutions (under suitable regularity conditions on the coefficients and source terms) to the linear first order symmetric hyperbolic equations of the form \eqref{eq: 2new_div-type-eqs}.

The real strength of the proposed method emanates from the freedom we have in choosing the applied $1+2$ decomposition. As we saw no matter how the foliation, determined by a smooth real function $\rho: \Sigma \rightarrow \mathbb{R}$, and the flow vector field $\rho^i$ are chosen the pertinent $\rho$ coordinate will always play the role of time in the pertinent evolutionary form of the constraints. In order to provide some evidences concerning the capabilities and some of the prosperous features of the proposed method the basic steps of initializing the time evolution of a pair of interacting point charges were also outlined. This simple example should also provide a clear manifestation of the consent which always comes along with the use of the proposed evolutionary form of the constraints in electrodynamics.  

\section*{Acknowledgments}

The author is deeply indebted to Jeff Winicour for his careful reading and for a number of helpful comments and suggestions.This project was supported by the POLONEZ programme of the National Science Centre of Poland which has received funding from the European Union`s Horizon 2020 research and the innovation programme under the Marie Sk{\l}odowska-Curie grant agreement No.~665778. 
	
	\includegraphics[scale=0.42, height = 42pt]{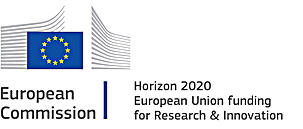}



\end{document}